\documentstyle[graphicx,referee]{mn}

\title{Meteors And Showers A Millennium Ago}
\author[Ahn]{Sang-Hyeon Ahn 
\\
Korea Institute for Advanced Study, 207-43 Cheongyangri-dong, Dongdaemun-gu, Seoul 130-722, Korea}

\date{2003 Feb}

\begin{document}

\maketitle

\begin{abstract}
Meteors can be classified into either sporadic meteors or 
showery meteors. We compile the meteor records in the astronomical archives 
in the Chronicle of the Koryo dynasty (918-1392), and investigate 
the spatial distribution of meteor streams along the orbit of 
the Earth from the 10th century to the 14th century.
We see that meteors from meteor streams signalize themselves over 
noisy sporadic meteors, and that the seasonal activity of sporadic meteors
was apparently regular. We discover the presence of a few 
meteor streams by analysing about 700 meteors in the Koryo period.
We also compile the records of meteor showers and storms
in the chronicles of Korea, Japan, China, Arab, and Europe,
and compare their appearance dates with those of showers 
obtained above, as well as with the modern observations. 
We confirm that the three sets of data are in agreement with each other. 
The representative meteor showers are the Perseids, the Leonids, and 
the $\eta$-Aquarids/Orionids pair formed by Halley's comet.
The other weak or relic meteor streams are also observable but uncertain. 
Hence, we witness the regularity of meteor activity, which is seen
to persist for a millennium.
\end{abstract}

\section{Introduction}
Comets scatter their evaporated particles around their orbit,
and these materials eventually fill the cometary orbit 
due to the dispersion of momenta. The belts of such remnants are 
called meteor streams. When the Earth passes through meteor streams,
meteor showers are seen on the sky.
In particuler, just before and after a comet passes through its perihelion,
its meteor stream is greatly enriched and the meteor shower 
looks almost like a firework. This is often described as a meteor storm.
After a long time, cometary materials are dispersed 
into the interplanetary space.
If the dispersed meteoroids enter the Earth's atmosphere, we may see
sporadic meteors. Presently the rate of major meteor showers is 
approximately one meteor shower per month. Meteors are observed and 
recorded by astronomers around the world both
optically and in radio. Regular pattern of annual meteor activity is 
well illustrated in Figure 6 and Table 2 of Yrj\"ol\"a and Jenniskens (1998). 

Due to the advancement of observational techniques, weak meteor showers 
have become an attractive topic in this field.
Hughes (1990) discovered that the apparently random sporadic meteors 
are not so random, 
and that about 20\% of meteors during a non-shower day belong to weak showers.
He also suggested that the flux of sporadic meteors 
can vary by about 30-40\% over $10^4-10^5$ years
due to the statistical variations in the numbers of massive short period
comets. 
Although modern astronomy provides precise and sensitive 
techniques for massive observation of meteors and showers, 
they cannot provide the data that can show the long-term variations
of meteor activity over more than a few hundred years.
Hence, it is worth discovering whether we can investigate
the long-term variation of meteors and showers 
by analysing the astronomical records of ancient people.

Another important factor that determines the meteor flux,
either sporadic or showery, is the distance between the Earth's orbit
and the nodal points of the cometary orbits.
The orbits of comets are affected 
by gravitational perturbations due to major objects in the Solar system.
The non-gravitational forces
are also strong sources of orbital changes, but it is very hard to
calculate the effects even in the modern astronomy.
Hence, the ancient astronomical records in chronicles can be useful 
to understand the long-term effects of these forces on orbital displacement.

In this study we will investigate the meteor records written in the 
Koryo dynasty (918-1392) which is one of the dynasties in the Korean history.
In Section 2 we explain which archives are authentic and reliable 
and how we compile the data. The analysis method is explained in Section 3,
and the results are followed in the next Section.
In the final Section we discuss our results.

\section{Korean Records}

Korean astronomical records written in history books have several merits.
Firstly, the duration of a Korean dynasty was typically around 500 years,
and so the astronomers in the Koryo dynasty could observe
the astronomical phenomena continuously. 
Moreover ancient Korean dynasties established their own Royal
Observatories, and hired a number of professional astronomers.
The astronomers recorded observational data in official documents,
which were systematically compiled into Royal history books 
after kings' death. Kings could not see the historical records of
his own, as well as those of preceding kings.
Thus the historians did not need to forge or discard the astronomical
phenomena which were usually interpreted as bad omens against kings.
In China, contrary to Korea, kings could consult their own history 
books and documents during their lives.
Japanese records are of limited use because they are few in number.
Nha (1997) investigated the records of solar and lunar eclipses
in {\it The Chronicle of the Koryo dynasty} ({\it Koryosa}),
and confirmed the regularity of the astronomical data during 
the Koryo dynasty (918-1392). Therefore, although other factors might 
also cause recording bias, the above factors support 
the reliability of the Koryo data.

\begin{table*}
 \centering
  \caption
{The official chronicles in the Korean history.}
  \begin{tabular}{@{}lll@{}}
    dynasty    &  duration     & history books \\
    [10pt]
Old Choson     & 2333 B.C. - 108 B.C. & The Ancient Chronicle of the Old Choson dynasty ({\it Huandan kogi}) \\
Three Kingdoms & 57 B.C. - 668 A.D.   & The Chronicle of the Three Kingdoms ({\it Sam-Guk-Sa(-Gi)}) \\
Parhae         & 698-926 A.D.         & -  \\
Unified Shilla & 676-935 A.D.         & The Chronicle of the Three Kingdoms ({\it Sam-Guk-Sa(-Gi)})  \\
Koryo          & 918-1392 A.D.        & The Chronicle of the Koryo dynasty ({\it Koryosa}) \\
               &                      & The Simplified Chronicle of the Koryo dynasty ({\it Koryosa-Joryo}) \\
Choson         & 1392-1910 A.D.       & The Chronicle of the Choson dynasty \\
\end{tabular}
\end{table*}

The majority of astronomical records in the Korean history are preserved 
in a few major history books published by the historians of 
the successive dynasties. Some history books were also published privately. 
A number of the astronomical records of the Old Choson dynasty are 
remained in a book titled by {\it The Ancient Chronicle of the Old Choson
dynasty}. However, this book was discovered in modern times,
and unfortunately historians and bibliographers suspect
that it may be a forged one. 
The astronomical records in the era of the Three Kingdoms 
(57 B.C. - 668 A.D.) and the Unified Shilla dynasty (668-935) are written 
in {\it The Chronicle of the Three Kingdoms} 
({\it Sam-Guk-Sa}\footnote{The name 
of this book is widely known as {\it Sam-Guk-Sa-Ki.}}) (Kim et al. 1145).
Almost no astronomical records of the Parhae dynasty (698-926) are left
due to the lack of its own history books.
The astronomical records during the Koryo dynasty (918-1392) are preserved 
in {\it The Chronicle of the Koryo dynasty} ({\it Koryosa}) (Kim et al. 1451)
and {\it The Simplified Chronicle of the Koryo dynasty} ({\it Koryosa-Joryo})
(Kim et al. 1452).  
The information during the Choson dynasty (1392-1910)
are written in {\it The Chronicle of the Choson dynasty}
({\it Choson-Wang-Jo-Shillok}). These are surely vast sources of 
astronomical data for modern astronomy. 
The Korean records have been introduced to the international society 
mainly by foreign scholars (e.g. Imoto \& Hasegawa 1958). 
However, they did not refer to the original historical books
such as those in Table 1. They merely used digested 
versions of Korean history books, such as {\it Mun-Heon-Bi-Go} 
and {\it Yol-Sung-Shillok}, which have a number of ambiguous or 
wrong dates and misleading expressions for astronomical events.

In this paper we concentrate on the meteor records of the Koryo dynasty.
The records have never been catalogued after thorough examination
on errors in their dates, and so we first made the catalogue after correcting
several erroneous dates.
We use the data in the authentic history book or {\it Koryosa}.
The records of meteor storms in the Korean history books 
are also surveyed. The detailed procedure of compilation was published 
in another Korean journal (Ahn et al. 2002).

A typical record of meteor reveals a few kinds of information such as 
date, time, starting and ending points, brightness, colour, and sound.
The dates were represented by the calendar system used at that time,
and they can be converted into Julian dates by utilizing the calendar
conversion table established precisely 
before this work (Han 1996, Shim et al. 1999).
In {\it Koryosa} there are 729 meteor records in total. 
One of them has no date, eight of them have ambiguity 
in calendar system, and fifteen of them were described as precursors of 
a meteor storm. These fifteen records are also excluded in our analysis, 
because they may bias the result.
Hence, we have 705 meteors which can be analysed. 

There appears no meteor records in the 10th century, 
to say nothing of other kinds of astronomical records.
The extreme rarity of historical data during this century was caused
by the invasion of the Kitanese army in 1011 A.D.. When the army occupied 
the capital city of Koryo, they fired the Korean documents and books of 
former ages. However, the Korean records from the 11th century 
to the 14th century show largely no discontinuity of records. 
On the other hand, the Chinese records show a bimodal distribution,
in that there are few records in the reign of 
the Mongolian Empire (Hasegawa 1992).

\section{Method}

Events of sporadic meteors resemble random noise, 
while meteors belonging to meteor showers can be considered as a signal.
The meteors caused by a meteor stream are likely to fall 
at nearly the same time of year, while sporadic meteors fall randomly.
Hence, any concentration in appearance dates of meteors in {\it Koryosa}
can be thought of as a sign of a meteor shower. 

For meteor showers, astronomers usually rely 
on the solar ecliptic longitude ($\lambda_\odot$)
which is defined by the angle in degrees along the ecliptic 
from the Vernal equinox to the position of the Earth.
Evidently the ecliptical coordinate system 
(equivalently the Julian and the Gregorian calendar systems) is not
convenient to compare two events whose interval is as large as thousand
years. That is because the origin of the ecliptical longitude
or the Vernal equinox varies with time due to the precession of
the Earth's rotation axis.

In this paper we first locate meteor showers a millennium ago,
and then compare their appearance dates with the modern observations 
to identify the meteor showers. In order to do so, we define 
the number of days after the perihelion passage time of the year, 
which is denoted by $\Lambda$.
In other words it is equivalent to relocate the origin of the ecliptic
coordinate from the Vernal equinox to the perihelion.
Since the orbit of the Earth is more stable than the Earth's rotation,
$\Lambda$ is more adequate for our purpose.
The perihelion passage time of the year is 
calculated by using the method in Meeus (1991). 

\section{Results}

\subsection{Seasonal Variation of Sporadic Meteors}

The results are shown in Figure 1, where peaks appear 
over a fluctuating background of sporadic meteors.
These prominent peaks represents the major meteor showers in the Koryo period.
According to Yrj\"ol\"a \& Jenniskens (1998),
the level of sporadic activity can be expressed by a functional form:

\begin{equation}
N(\lambda_\odot) = \langle N_{spo} \rangle
- \Delta N_{spo} \cos (\lambda_\odot),
\end{equation}
where $\langle N_{spo} \rangle$ is mean daily sporadic hourly count 
at the Summer solstice on June 21 and $\Delta N_{spo}$ is 
the yearly amplitude due to the seasonal variation. Here $\lambda_\odot$ 
is the solar ecliptical longitude. The Koryo data are too sparse to
give us meaningful values for $\langle N_{spo} \rangle$ and
$\Delta N_{spo}$. We need more detailed description and additional data 
such as weather conditions to compensate the sparseness of data,
but there are few such detailed data in history books of the Koryo era. 
Hence, we will simply check the sinusoidal variation of seasonal activity 
of sporadic meteors in this study.
Thus, we make a simple eye-fitting to the sporadic flux by Eq.(1).
Here we exclude the prominent peaks over the fluctuating background.
Data around $\Lambda = 200$ are also excluded, because they
can bias results. This is a season of {\it Jangma}, during which it
rains continuously, being related with the East Asian Monsoon.
However, the fitting is merely qualitative and can be thought of
as an adjoint line in Figure 1. 

We can see in the figure that the sporadic fluctuations show rough 
agreement with the fitting curve. This general trend that 
the flux is small in spring and large in autumn, is 
caused by the fact that the Earth's rotation axis is inclined and 
so altitudes of radiant points are higher in autumn than in spring. 
Our result supports the idea that the meteor data of 
the Koryo dynasty is faithfully representing the natural phenomena.

\subsection{Prominent Meteor Showers}

Now let us turn our attention to the major meteor showers which
appear as peaks in Figure 1. The most prominent
peak appears on about $\Lambda=225$.
When we convert this date into the ecliptical longitude, we can see 
that it corresponds to $139^\circ \le \lambda_\odot \le 142^\circ$.
(In this paper, the solar longitudes of meteor showers are 
expressed for B1950.)
According to the table given by Yrj\"ol\"a \& Jenniskens (1998),
this corresponds to the Perseids ($\lambda_\odot = 139^\circ$),
whose parent comet is 109P/Swift-Tuttle.
Since the orbit of this comet has a nearly normal inclination,
the gravitational perturbation of other objects in the solar system
is rather small, and consequently its orbit
is preserved for a long time.

The other prominent peaks appear in winter ($250 \le \Lambda \le 360$).
One of the most strong peaks 
appears in the period $297^\circ \le \Lambda \le 316^\circ$
which corresponds to $213^\circ \le \lambda_\odot \le 232^\circ$.
We can find two current major meteor streams in this period, 
which are the Taurids
($\lambda_\odot = 223^\circ$) and the Leonids. The parent comet of the 
Taurids is 2P/Encke, and its orbital inclination is merely $i=12^\circ.4$.
However, since its apheilon distance amount to
4.1 AU, the comet does not experience great gravitational 
perturbation of the Jupiter, so that its orbit is persistent over centuries
(Sitarski 1988). The orbital period of comet Encke is only 3.3 years,
which is easily short enough to be depleted in dust. Hence, the comet 
cannot supply lots of meteoroids to the meteor stream these days. 
However, historical studies indicate that activity of the Taurids during
the 11th century may have been comparable to today's Perseids
(Astapovic \& Ternteva 1968, Hasegawa 1992, Bone 1993).
However, it should be noted that the active durations of the Taurids 
and the Leonids overlapped in the period of Koryo, so that it is not easy 
to discriminate between them.

The present ecliptical longitude of the Leonids 
is $\lambda_\odot \approx 235^\circ$, which implies its position 
on the ecliptic has shifted over the years. In fact, since the precession of
the nodal points of the parent comet, 55P/Tempel-Tuttle, is 1.6 day 
per century (Mason 1995), we can estimate the ecliptical longitude of 
the Leonids during 10-14th century to be 
about $220^\circ \le \lambda_\odot \le 225^\circ$, which is 
in good agreement with the peak seen above at $250 \le \Lambda \le 360$. 
Contrary to either 109P/Swift-Tuttle or 2P/Encke,
the parent comet of the Leonids, 55P/Tempel-Tuttle, has
an orbit that is close to the ecliptic, so that the stronger gravitational
perturbation makes the orbit precess.

In spring there is another shower at $\Lambda=126$ or
$\lambda_\odot = 48^\circ$. The signal is very marginal, but its ecliptical
longitude corresponds to the $\eta-$Aquarids, whose parent comet is
Halley's comet. It is well-known that this shower has a pair shower,
the Orionids at $\lambda_\odot=208^\circ$.
We can find the weak feature of the Orionids at $\Lambda=289$
or $\lambda_\odot=205^\circ$. Since the bin size of the horizontal axis 
in the figure is as large as 4 days, we can say that this small peak 
is likely to be the Orionids. When we compare 
the annual meteor activity in the Figure 6 of Yrj\"ol\"a \& Jenniskens (1998),
we can see that both meteor showers are relatively weak in strength.
The signals are observed to be weak, but
the pairness of their appearance and the strong activity seen in
the data of the meteor storm for both showers indicate that
the $\eta-$Aquarids/Orionids were active in the Koryo period.

\subsection{Weak Showers}

Another conspicuous peak is seen at $\Lambda = 270$, corresponding to
$\lambda_\odot = 186^\circ$. In the Table 2 of Yrj\"ol\"a \& 
Jenniskens (1998), we can find that the Sextantids has the coincident 
ecliptical longitude $\lambda_\odot = 186^\circ\pm2^\circ$.
However, its ZHR\footnote{Zenithal Hourly Rate}
is estimated to be merely 9 meteors per hour, which is very weak. 
However, we can not exclude the possibility that the Sextantids was 
stronger in the medieval era than now.

There appears a prominent peak at $\Lambda=335$.
Current major meteor showers observed in December
are rather variable in strength. Representative ones are the Bo\"otids
and the Geminids. Their present ZHR is as large as 110 and 140, respectively.
Although we see in Figure 1 a prominent peak
at $\Lambda=335$ or $\lambda_\odot=251^\circ$. 
This peak does not correspond to either the Bo\"otids or the Geminids.
According to Table 2 of Yrj\"ol\"a \& Jenniskens (1998),
this peak may correspond to the $\chi-$Orionids.

The peaks at $\Lambda=34$ and $\Lambda=95$ may correspond 
to the recent showers, the Feb-Draconids and the N-Virginids,
respectively. When considering another catalogue of meteor showers (Bone 1993),
these weak peaks are believed to be aged and depleted showers or 
relics of dead comets. However, the existence and the activity level 
of these weak showers in the Koryo period are still uncertain. 

\subsection{Temporal Variations}

In order to check the possibility of the temporal variation and 
confirm the validity of our results, we divide the data into two sets, 
around 1150 A.D.. The histogram is even more fluctuating 
because of the small amount of data. However, the seasonal
variation of the meteor flux and the conspicuous peaks such as 
the Perseids and the possible Leonids around $\Lambda=300$ can be 
seen in Figure 2. The dotted lines represent the annual variation of 
sporadic meteors observed recently.
This indicates that the meteor activity had persisted 
during the Koryo dynasty, and that the showers we discovered are 
realistic and meaningful. This is also supported by other researches 
for the independent data set compiled out of the history books 
of the Sung and the Ming dynasties (Hasegawa 1992), and for
the Koryo data (Hasegawa 1998).
We will discuss this in Section 5.

\subsection{Meteor Storms}

Roughly speaking, the meteor storms written in chronicles form a 
subset of meteor showers, but they are splendid enough to be 
frequently recorded in ancient chronicles. 
We define the data sets as follows: (A) meteor showers found 
by analysing the data of sporadic meteors in {\it Koryosa},
(B) meteor showers and storms recorded in historical books listed below, 
(C) meteor showers observed recently.

It is much interesting if three data sets have some correlations.
Motivated by this fact, we check the consistency among these three 
sets of data. We collect the records of meteor showers and storms from 
chronicles of Korean, Japanese, Chinese, Arabian, and European countries
(Dall'olmo 1978, Kanda 1935, Rada \& Stephenson 1992, Hasegawa 1996, 
Mason 1995, Imoto \& Hasegawa 1958, Beijing Observatory 1988).
We collect the Korean records of meteor showers with a thorough
survey of the Korean archives, and the catalogue will be 
published elsewhere. 
After appearance dates of all the data being converted into the same 
time coordinate or $\Lambda$, we check the conformity in appearance 
dates between the showers and the storms. We show in Figure 3 
their appearance dates in year.
The broad stripes that are yellow, blue, and red in colour 
represent the active duration of 
each current meteor shower denoted on the left of the box. 
The narrow lines represent the durations of maximal activity for each
current meteor shower. The symbols represent the meteor storms reported 
by civilizations, and each meaning can be consulted in the figure caption.

First of all, it is remarkable in the figure that the current maxima 
of the $\eta-$Aquarids and the Orionids, remnants 
of Halley's comet, show an exact conformity to
those of the historical meteor storms.
The Perseids also shows similar agreement, and the appearance date 
seem to have been a few days later than now. That is to say, its $\Lambda$
was a few days larger than now.

The Leonids displays its firework in a relatively short duration,
and their displaying date has been delayed year by year 
at a rate of 1.6 days per century (Mason 1995). 
According to the compilation of Mason (1995),
the earliest record of the Leonid storm is traced 
back to 902 A.D. in Egypt.
However, the Orionids overlapped with the Leonids around the 9th century.
Hence, even if there are any record of the Leonids in any chronicle, 
it is uncertain whether it belongs to the Leonids or the Orionids.

\section{Discussion}

We have investigated the sporadic meteors and showers in the Koryo dynasty
(918-1392). We found that the seasonal activity of sporadic meteors shows 
rough agreement with the modern observation. We also found that there were 
a few prominent meteor showers and storms: the Perseids, the Leonids, 
and the possibly $\eta$-Aquarids/Orionids pair. 
There were also weak and aged meteor showers. 
The meteor showers known from the Koryo data are in good agreement with
the meteor storms compiled from the world-wide chronicles,
as well as with the modern observation. 

Our results can be supported by analyses of independent data 
in another historical archives such as the Sung dynasty (960-1279)
and the Ming dynasty (1368-1644).
Hasegawa (1992) investigated the Chinese 
and the Japanese records of meteors. The number of the Chinese records 
during the period of the Koryo dynasty (918-1392) amounts to 1638, 
while that of the Koryo records is 729. 
The Chinese meteor records increase sharply in number around 1070 A.D.
and 1430 A.D., just at the beginnings of the Sung and the Ming dynasties.
This may be attributed to the changes in the principles or 
the methods of the meteor observation in the dynasties (Hasegawa 1992).
Also there are few meteor records during the Yuan dynasty (1271-1368)
of the Mongol empire.
On the other hand, the temporal distribution of the meteor records 
in the Koryo dynasty has a broad peak around the 12-13th centuries.
Hence two sets of data are statistically independent and complementary 
with each other, because both the observers and the characteristic 
periods are different.

Hasegawa (1992) investigated the monthly variations of meteors 
recorded in China during 1-1000 A.D. and all centuries after 1001 A.D.,
which was shown in Figure 3 of his paper. He discovered two conspicuous 
maxima in July-August and November-December.
He thought that these two maxima are in agreement with recent observations,
and insisted that this supports the reliability of the Chinese records. 
He concluded that the July-August peak is the Perseids, and that 
the November-December peak is the Taurids.

Hasegawa (1998) also applied his work to the Koryo meteor records,
and concluded that the meteor records of the Koryo dynasty show
the same pattern in the variations of meteor flux as seen 
in the Chinese records. However, unfortunately, he only inspected
the monthly variations in both papers, and so the time-resolution
is not fine enough to see the individual meteor showers.
The temporal resolution in our study is about five days.
As a result, we can see that the Perseids existed a thousand years ago, 
and that the Hasegawa's November-December shower is not likely 
to be the Taurids, but the Leonids, because of its appearance date
and the coincidence with the records of the meteor storms. 
At any rate, Hasegawa's study supports our conclusions.

The meteor records of the Choson dynasty (1392-1910) were written
in the Chronicle of the Choson dynasty. The total number of records amounts
to about 3500. We have been carrying out the similar researches on
the Choson data, and the analysis for the part of records 
in the Choson dynasty have been finished. 
The results support our conclusion in this paper.
The results will be published elsewhere.

We conclude that the major meteor showers and storms a millennium ago
might be caused by the same short-period comets to the current ones,
and their activities were similar to those of present showers.
It is interesting that although everyday meteors
seem to be sporadic and random, there is a regularity, 
which lasts for at least a millennium.

\section*{acknowledgements}
It was a great pleasure for me to make a catalogue of the meteor records 
during the Koryo dynasty with Mr. Bae, Mr. Jeong, and Ms. Cho of
Yonsei University, Seoul, Korea. The author also thanks to Dr. Julian Lee 
for his useful criticism and careful reading this manuscript.

\begin{figure*}
\centering
\includegraphics[width=18cm,angle=0]{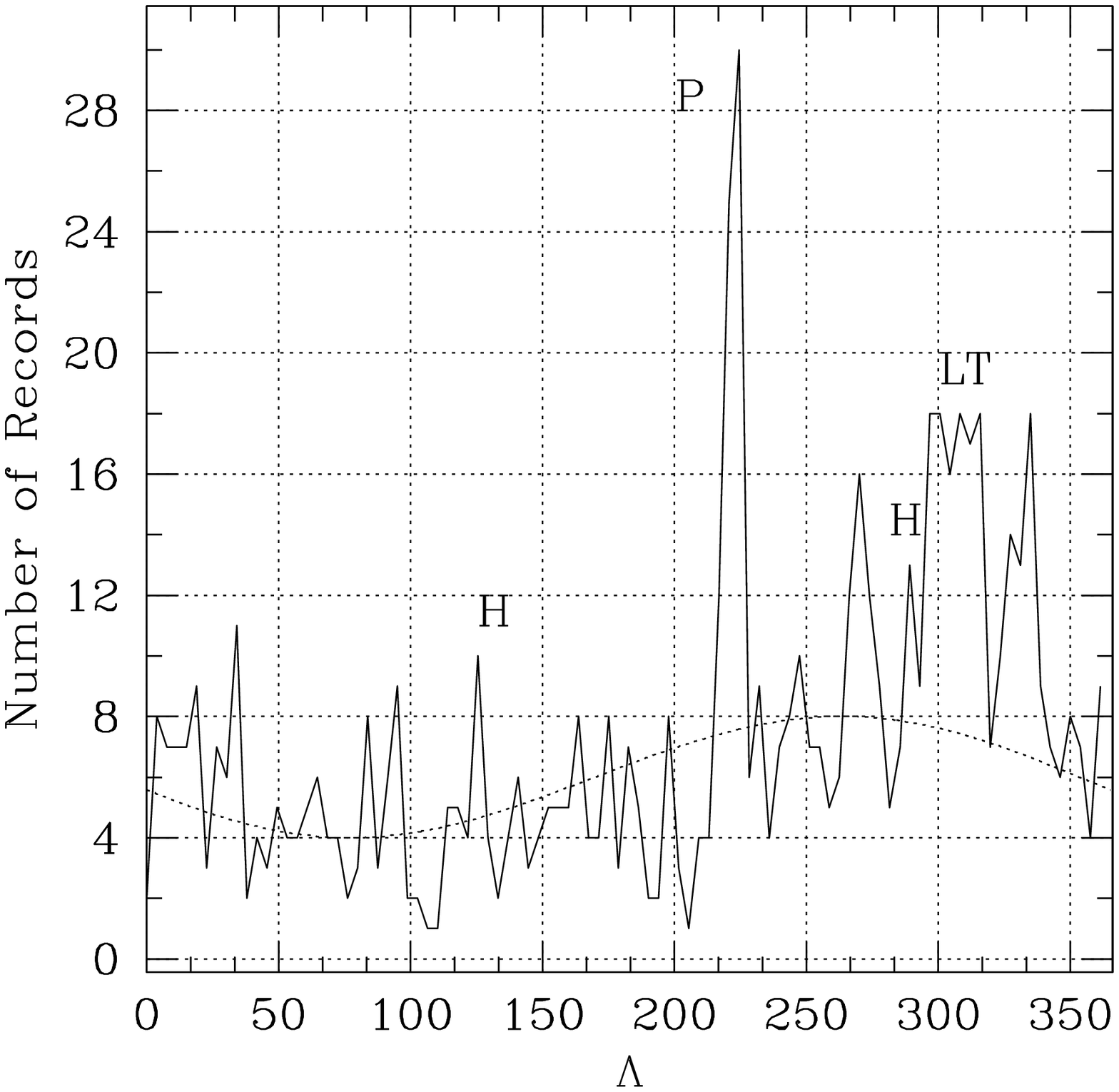}
\caption{
\small
Annual meteor activity extracted from the meteor data in the
astronomical archives of the Koryo dynasty (918-1392).
In the horizontal axis, $\Lambda$ is days from the perihelion 
passage time of that year to the apparition of meteors.
The sinusoidal dotted line stands for the fitting curve of 
the recently observed annual variation of sporadic meteors.
The peak denoted by P is the Perseids, LT means the Leonids
and the Taurids mixture, and two H's mean the $\eta-$Aquarids 
and the Orionids, whose parents is Halley's comet.}
\end{figure*}

\begin{figure*}
\centering
\includegraphics[width=18cm,angle=0]{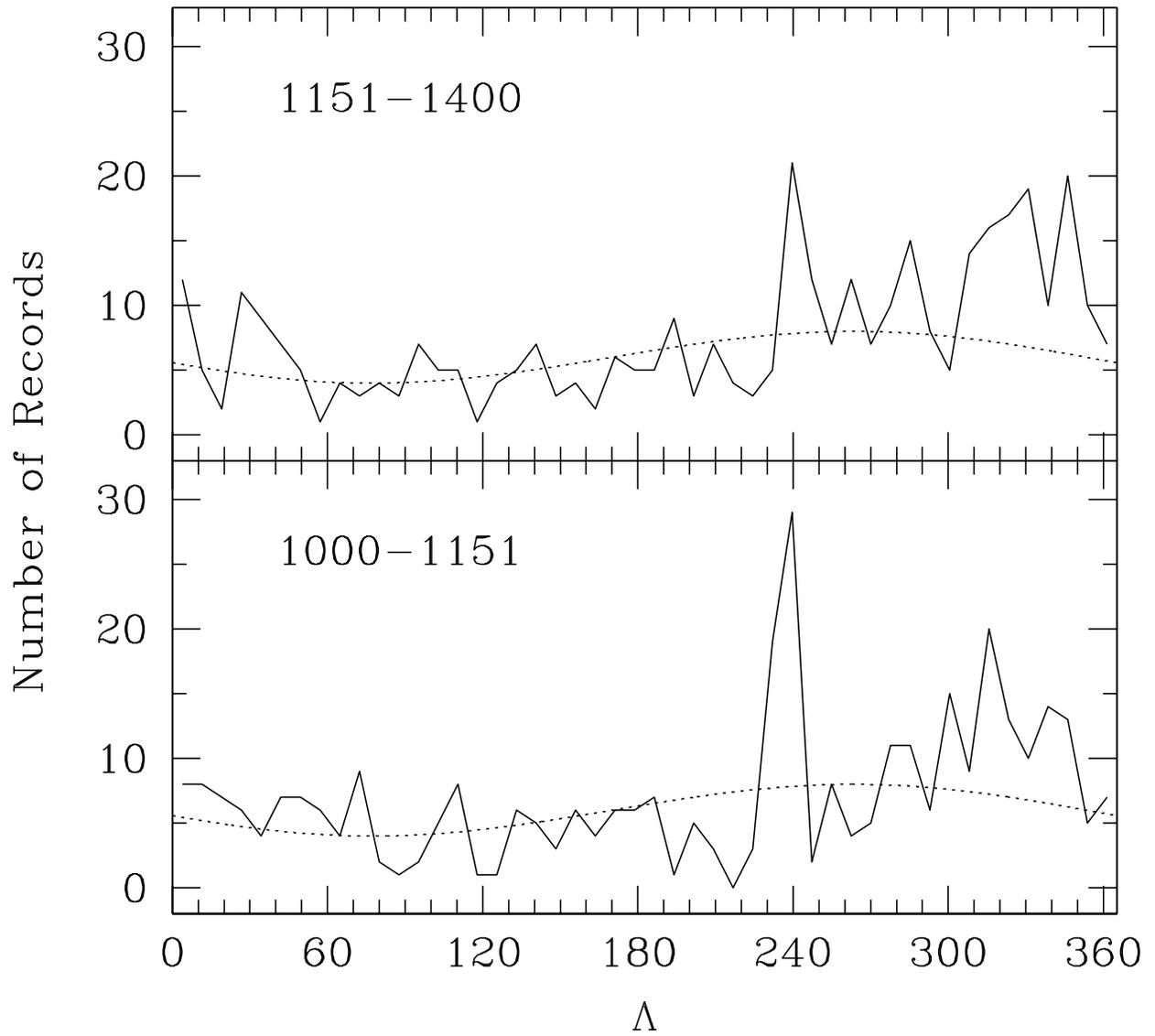}
\caption{
\small
Annual meteor activity extracted from the meteor data in the
astronomical archives of the Koryo dynasty (918-1392).
In the horizontal axis, $\Lambda$ is days from the perihelion
passage time of that year to the apparition of meteors.
Here we divided the data into two sets around 1150 A.D..
The sinusoidal dotted lines stand for the eye-fitting curve of 
the recent seasonal variation of sporadic meteors.}
\end{figure*}

\begin{figure*}
\centering
\includegraphics[width=18cm,angle=0]{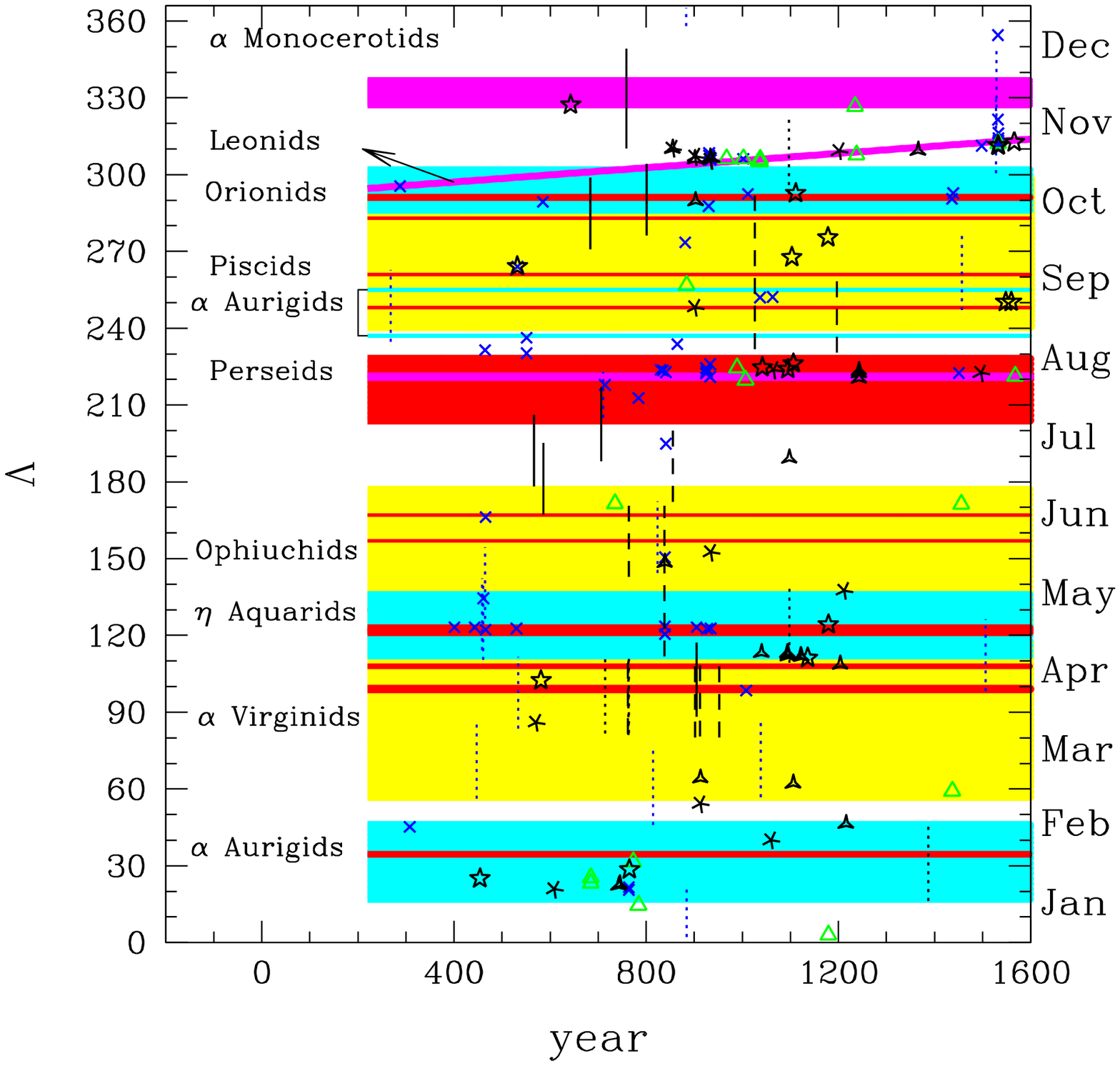}
\caption{
\small
Historical meteor showers drawn from the world wide 
data sources of east Asian, Arabian, and European countries.
The thick stars and the solid vertical bars represent the Korean data, 
the blue cross and the dotted vertical bars the Chineses data,
the thin stars and dashed vertical bars the Arabian data,
the triangular stars the Japanese data, and the green empty triangle
the European data. THe veritical bars represent the data whose dates
are given only upto month. The wide stripes 
represent the active duration, and narrow stripes in them 
stand for the maximum activity. The names of major showers are
written on the left side of the stripes. The vertical axis
is the number of days after the perihelion passage for each shower.}
\end{figure*}

\begin{thebibliography}{}
\bibitem{ahn03} Ahn S.-H., Bae H.-J., Cho H.-J., Jeong S.-W., 2002, Publication of Korean Astronomical Society, 17, 23
\bibitem{ast68} Astapovic I. S. \& Terenteva A. K., 1968, IAU Symposium no. 33, Dordrecht, D. Reidel, p.308 
\bibitem{bei88} Beijing Observatory, 1988, General Compilation of Chinese ancient astronomical records
\bibitem{bon93} Bone N., 1993, Meteors, (Sky Publishing Corporation: Cambridge Massachusetts)
\bibitem{dal78} Dall'olmo U., 1978, Journal of History of Astronomy, 9, 123
\bibitem{han96} Han B., 1996, Korean Calendar Conversion Table, (Daegu: Korea)
\bibitem{has92} Hasegawa I., 1992, Celestial Mechanics and Dynamical Astronomy, 54, 129
\bibitem{has96} Hasegawa I., 1996, QJRAS, 37, 75-78
\bibitem{has98} Hasegawa I., 1998, in {\it Dynamics of Comets and Asteroids and their Role in Earth History}, Proceedings of a Workshop held at the Dynic Astropark 'Ten-Kyu-Kan', 14-18 August, 1997. Edited by Shin Yabushita, and Jacques Henrard. (Kluwer Academic Publishers), 279
\bibitem{hug90} Hughes D., 1990, MNRAS, 245, 198
\bibitem{imo58} Imoto S. \& Hasegawa I., 1958, Smithsonian Contributions to Astrophysics, 2, 131
\bibitem{jen01} Jenninskens P., 2001, Journal of the International Meteor Organization, 29, no. 5, pp.165-175
\bibitem{kan35} Kanda S., 1935, Japanese historical records of celestial phenomena, Tokyo
\bibitem{sam} Kim B.-S. et al., 1145, The Chronicle of the Three Kingdoms ({\it Sam-Guk-Sa(-Gi)}), 1145, Koryo
\bibitem{koryo} Kim J.-S. et al., 1451, The Chronicle of of the Koryo dynasty ({\it Koryosa}), Choson
\bibitem{jeol} Kim J.-S. et al., 1452, The Simplified Chronicle of the Koryo dynasty ({\it Koryosa-Jeoryo}), Choson
\bibitem{mas95} Mason, J. W., 1995, J. of British Astron. Assoc., 105, 219
\bibitem{mee91} Meeus J., 1991, Astronomical algorithms, (Willman-Bell Inc.:Virginia Richmond)
\bibitem{nha97} Nha I.-S., 1997, The Journal of Korean Studies, 96, 1
\bibitem{rad92} Rada, W. S. \& Stephenson F. R., 1992, QJRAS, 33, 5
\bibitem{wang} Royal Historians, The Chronicle of the Choson dynasty ({\it Choson-Wangjo-Shillok}), Choson
\bibitem{shim99} Shim K.-J., Ahn Y.-S., Han B.-S., Yang H.-J., Song D.-J., 1999, Calendar Table of the Koryo dynasty, (Korea Astronomy Observatory: Daejeon Korea)
\bibitem{sit68} Sitarski G., 1988, Acta Astronomica, 38, 269
\bibitem{yrj98} Yrj\"ol\"a I. \& Jenniskens P., 1998, A\&A, 330, 739
\end{thebibliography}
\end{document}